\documentclass[%
reprint,
amsmath,amssymb,
aps,
pra,
]{revtex4-1}

\usepackage{graphicx}
\usepackage{dcolumn}
\usepackage{bm}

\usepackage[colorlinks,linkcolor=blue,anchorcolor=blue,citecolor=blue]{hyperref}
\begin{document}
	
	\preprint{APS/123-QED}
	
	\title{Practical security analysis of a continuous-variable quantum random-number generator with a noisy local oscillator
	}
	
	\author{Weinan Huang$^{1}$}
	\author{Yichen Zhang$^{1}$}%
	\email{Correspondence: zhangyc@bupt.edu.cn.}
	\author{Ziyong Zheng$^{1}$}%
	\author{Yang Li$^{2}$ }%
	\author{Bingjie Xu$^{2}$}%
	\author{Song Yu$^{1}$}%
	\affiliation {%
		$^{1}$ State Key Laboratory of Information Photonics and Optical Communications, Beijing University of Posts and Telecommunications, Beijing, 100876, China \\
		$^{2}$ Science and Technology on Security Communication Laboratory, Institute of Southwestern Communication, Chengdu, 610041, China
	}%
	
	
	
	\date{\today}
	
	\begin{abstract}
		A quantum random-number generator (QRNG) can theoretically generate unpredictable random numbers with perfect devices and is an ideal and secure source of random numbers for cryptography. However, the practical implementations always contain imperfections, which will greatly influence the randomness of the final output and even open loopholes to eavesdroppers. Recently, Thewes et al. experimentally demonstrated a continuous-variable eavesdropping attack, based on heterodyne detection, on a trusted continuous-variable QRNG in Phys. Rev. A 100, 052318 (2019), yet like in many other practical continuous-variable QRNG studies, they always supposed the local oscillator was stable and ignored its fluctuation which might lead to security threats such as wavelength attack. In this work, based on the theory of the conditional min-entropy, imperfections of the practical security of continuous-variable QRNGs are systematically analyzed, especially the local oscillator fluctuation under imbalanced homodyne detection. Experiments of a practical QRNG based on vacuum fluctuation are demonstrated to show the influence of local oscillator fluctuation on the total measurement noise variances and the practical conditional min-entropy with beam splitters of different transmittances. Moreover, a local oscillator monitoring method is proposed for the practical continuous-variable QRNG, which can be used to calibrate the practical conditional min-entropy.
	\end{abstract}
	
	
	\maketitle
	
	
	\section{\label{set1}Introduction}
	
	Random numbers are paramount ingredients for varied applications from cryptography to numerical simulation, even gaming and lotteries. Traditional random sources of mathematical algorithms and classical physical processes will not be suitable choices to generate true random numbers due to their intrinsic determinacy~\cite{Nisan88,Petrie00,Stojanovski01}. Thanks to quantum mechanics, a quantum random-number generator (QRNG)~\cite{Ma16,Miguel17}, which supports the generation of unpredictable random numbers from the intrinsic uncertainty of quantum processes,  becomes the most appealing methods for random number generation. The existing QRNG protocols include discrete protocols based on photon arrival time~\cite{Ma05,Wahl11,Nie14}, photon branching path~\cite{Stefanov00,Jennewein00} and photon number distribution~\cite{Wei09, Ren11, Applegate15}, and continuous protocols based on phase noise~\cite{Guo10,Qi10,Xu12}, amplified spontaneous emission noise~\cite{Williams10,Wei12} and vacuum fluctuation~\cite{Christian10, Symul11, Haw15, Raffaelli18, Zheng19}. The continuous-variable (CV) QRNG based on measuring vacuum fluctuation has the advantage of simple implementation and high-speed potential, which has attracted much attention and undergone huge development since it was proposed~\cite{Christian10}, such as a generation speed of Gbps~\cite{Zhang16, Zheng19} and on-chip integration~\cite{Raffaelli18}.
	
	Real random numbers can be obtained by QRNGs constructed by ideal devices. To deal with the security problems caused by the practical devices, the most extreme protocol is the device-independent protocols~\cite{Pironio10, Liu18}. Although it can eliminate the influences of all non-ideal factors, the rate of random-number generation is very slow. Improved protocols are semi-device-independent protocols, which include source-device-independent protocols~\cite{Cao16,Xu17,Marangon17, Avesani18,Smith19,zhang2020finitesize} and measurement-device-independent protocols~\cite{Cao15,Nie16}. Semi-device-independent protocols relax the requirement for some devices by fully trusting other devices. Although the generation speed has been greatly improved, it still needs to trust some of the devices completely.
	
	\begin{figure*}[t]
		\includegraphics[width=18cm]{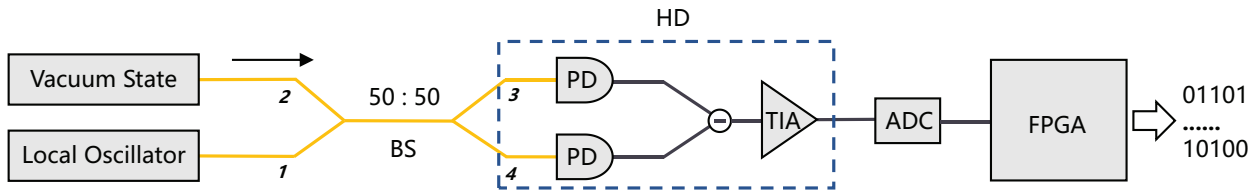}
		\caption{\label{setup_basic}Schematic of vacuum-based CV-QRNG. The LO interferes with the vacuum state in the symmetrical fiber BS. Then two beams are output with the same power and detected by the DC-coupling HD (shown in the dashed box) for photoelectric conversion, and the output voltages are sampled and quantized by the ADC to generate random bits. The effects of electronic noise on these random bits are removed by postprocessing operations by the FPGA. The unpredictable random numbers are finally obtained.
			(BS, beam splitter; LO, local oscillator; DC, direct current; HD, homodyne detector; PD, photodiode; TIA, trans-impedance amplifier; ADC, analog-to-digital converter; FPGA, field-programmable gate array).}
	\end{figure*}
	
	Unlike the previous protocols, the practical protocols characterize devices in the practical system and evade the impacts of various non-ideal factors utilizing calibration, monitoring, and other measures, so as to ensure the system's practical security under the premise of achieving high speed. As a practical continuous protocol, in addition to the improvement of the generation rate, the practical security of the vacuum-based CV-QRNG is also a major issue that must be considered. Some remarkable works focus on postprocessing methods with quantum-proof random-number extractors~\cite{Ma13}, the impacts of the sampling range and the accuracy of the analog-to-digital converter (ADC)~\cite{Haw15}, and the  security of the postprocessing method~\cite{Chen19}. In previous works on the practical security of the vacuum-based CV-QRNG, it is generally assumed that the local oscillator (LO) is from an ideal laser source, that is, there is no fluctuation in power, and the homodyne detection is balanced. The pity is that there is LO fluctuation in a practical system, and the homodyne detection is usually imperfect. Thus, if the LO fluctuation is ignored, one will overestimate the min-entropy, and it will be a loophole exploited by the eavesdropper. However, there is still a lack of study on the effect of the LO fluctuation on the practical security of the vacuum-based CV-QRNG under imbalanced homodyne detection, which will open a loophole for eavesdroppers.
	
	In this work, based on the theory of the conditional min-entropy, we systematically analyze the effects resulting from the defects of the practical devices, such as the LO fluctuation, the imbalance of the beam splitter, and the imperfect conversion efficiencies of photodiodes (PDs), on the practical security of the vacuum-based CV-QRNG. A basic variance model and a practical min-entropy model are developed. From the models, relations between the transmittance of the beam splitter and the variances, including LO fluctuation variance, quantum variance and total measurement variance, are inferred. To confirm these inferences, experiments are demonstrated in a practical vacuum-based CV-QRNG system with LO monitoring as beam splitters of different transmittances are used. Last but not least, the practical conditional min-entropy of a normal practical vacuum-based CV-QRNG is calibrated by considering the LO fluctuation and its generation speed is calculated.
	
	The organization of this paper is as follows. In Sec.~\ref{set2}, we review the basic principles of vacuum-based CV-QRNG, introducing a vacuum-based CV-QRNG protocol and a conditional min-entropy. In Sec.~\ref{set3}, we analyze the non-negligible impacts of LO fluctuation under imbalanced homodyne detection in detail and build a practical conditional min-entropy model. In Sec.~\ref{set4}, we demonstrate experiments in a practical vacuum-based CV-QRNG system with beam splitters of different transmittances and LO monitoring to verify the practical conditional min-entropy model. Furthermore, we calibrate the practical conditional min-entropy by considering the LO fluctuation and calculate the generation rate. In Sec.~\ref{set5}, we conclude the work that has been done in this paper and discuss possible attacks in the practical vacuum-based CV-QRNG system.
	
	\section{\label{set2}Basic Principles of Vacuum-based CV-QRNG}
	In this section, we will review the basic principles of a vacuum-based CV-QRNG, including the introduction of vacuum-based CV-QRNG  protocols and the conditional min-entropy.

	\subsection{\label{set2.1}Optical quantum random-number generator based on vacuum fluctuation}
	A vacuum-based CV-QRNG can generate random numbers at high speed by sampling the quantum randomness of vacuum fluctuation. Using the homodyne detection technique to measure the amplitude quadrature of the vacuum state, the measurement result is a distribution meeting a Gaussian probability function.
	Figure~\ref{setup_basic} shows the experimental setup of a vacuum-based CV-QRNG. The continuous beam (LO) emitted by the laser interferes with the vacuum state in the symmetrical fiber beam splitter. Then two beams are output with the same power and detected by the balanced detector for photoelectric conversion, and the output voltages are sampled and quantized by a ADC to generate random bits. The (quantum) side information that an attacker could have gained from these random bits is removed by postprocessing operations. In the meantime, a uniform distribution of outcomes from the Gaussian raw distribution is obtained. Finally, the unpredictable random numbers are obtained. Generally, in order to strictly test the randomness of generated random numbers, NIST-STS~\cite{Rukhin10} and DIEHARDER~\cite{Brown16} test suits are mainly used.
	
	In fact, not all bits obtained from sampling are secure, so we resort to postprocessing operations also known as random extraction to get secure random bits. For determining the extraction ratio that denotes the secure numbers per sample, the practical conditional min-entropy of the system has to be calculated ahead of time.
	
	\subsection{\label{set2.2}Conditional Min-entropy}
	For variable $X$, which obeys the probability distribution $\mathrm{P}_{\mathrm{X}}\left(x_{i}\right)$, the min-entropy is defined in unit bits as~\cite{Dodis08, Konig09}
	\begin{equation}
	H_{\min }(x)=-\log _{2}\left[\max _{x_{i} \in X} \mathrm{P}_{\mathrm{X}}\left(x_{i}\right)\right].
	\label{eq1}
	\end{equation}
	
	It represents the uniform random bits that can be extracted under the maximum probability of the eavesdropper guessing $X$.
	\begin{figure}[t]
		\centering
		\includegraphics[width=9cm]{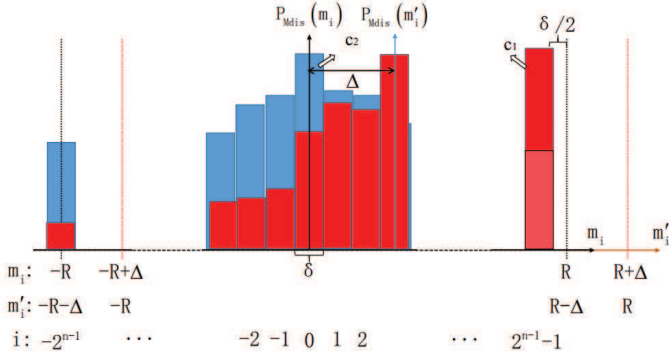}
		\caption{\label{c1_c2} Sampling model of the $n$-bit ADC~\cite{Haw15}, with the range $[-R+\delta / 2,R-3\delta / 2]$ and bin width $\delta=R / 2^{n}$. Offset $\Delta$ is introduced in a realistic scenario, thus the distribution of the original $m$ is now centered at offset $\Delta$ by another reference $m\prime$ and the lowest and highest bins are centered around $-R-\Delta$ and $R-\Delta-\delta$.}
	\end{figure}
	
	Suppose the electronic noise and the vacuum quantum noise are independent and identically distributed. And suppose the eavesdropper has the infinite computing power to completely master the classical noise $E$. That is to say, the classical noise $E$ can be acquired with infinitesimal sampling accuracy. Under this worst condition, the min-entropy of the measurement noise conditioned on the classical noise is~\cite{Haw15, Zheng19}
	\begin{equation}
	H_{\min }\left(M_{d i s} | E\right)=-\log _{2}\left[\max \left(c_{1}, c_{2}\right)\right],
	\label{eq2}
	\end{equation}
	where $M_{d i s}$ is a discrete measurement result satisfying the Gaussian distribution.
	$c_{1}=\frac{1}{2}\left[\operatorname{erf}\left(\frac{e_{\max }+\Delta_{\max }-R+3 \delta / 2}{\sqrt{2 \sigma_{Q}^{2}}}\right)+1\right]$ is the probability value of the boundary of the ADC sampling range. $e_{\max }$ and $\Delta_{\max }$ are respectively the maximum of the classical noise outcome $e$ and the maximum of the DC bias of the equipment. $R$ is the sampling range of the ADC, and the sampled signal is discretized into $2^{n}$ bins with bin width $\delta=R / 2^{n}$. $\sigma_{Q}^{2}$ is the variance of the vacuum quantum noise, and the probability value of ADC distribution in the middle of the sampling range is
	\begin{equation}
	c_{2}=\operatorname{erf}\left(\frac{\delta}{2 \sqrt{2 \sigma_{Q}^{2}}}\right)=2 / \sqrt{\pi} \int_{0}^{\frac{\delta}{2 \sqrt{2 \sigma_{Q}^{2}}}} e^{-t^{2}} d t,
	\label{eq3}
	\end{equation}
	The detailed sampling model of the ADC is depicted in Figure~\ref{c1_c2}.
	
	When $c_{1}>c_{2}$, it means that the eavesdropper can grasp more side information, and the vacuum-based CV-QRNG system will be suboptimal. Therefore, the conditional min-entropy model for analyzing the vacuum-based CV-QRNG can be simplified to analyze its probabilistic value by choosing the appropriate sampling range to make $c_{1} \leq c_{2}$. Equations (~\ref{eq2}) and (~\ref{eq3}) show that there is a positive correlation between $H_{\min }\left(M_{dis} | E\right)$  and $\sigma_{Q}^{2}$, so the analysis of the conditional min-entropy can be transformed into the analysis of $\sigma_{Q}^{2}$. In the practical vacuum-based CV-QRNG, the non-idealities of the devices will affect the conditional min-entropy, so it is necessary to analyze these defects of the practical devices.
	\begin{figure}[t]
		\centering
		\includegraphics[width=8.5cm]{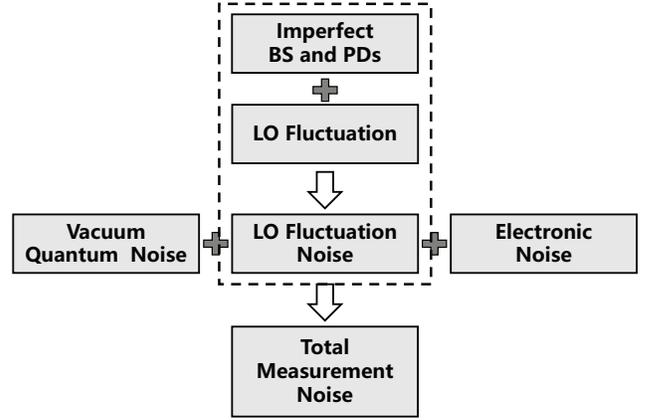}
		\caption{Variance model of the practical vacuum-based CV-QRNG. The existing protocols normally suppose the LO is ideal without fluctuation and homodyne detection is perfect; here we take the LO fluctuation and imperfect homodyne detection into account (shown in the dashed box).}
		\label{var_model}
	\end{figure}
	
	\section{\label{set3}Practical Conditional Min-entropy Model with LO Fluctuation}
	In this section, the non-negligible impacts of the LO fluctuation under imbalanced homodyne detection are analyzed in detail and a practical conditional min-entropy model is built.
	
	\subsection{\label{3.1}Basic practical variance model}
	Under ideal conditions, the LO fluctuation can be canceled out by the balanced detector. However, the non-idealities of practical devices, such as different quantum efficiencies of two PDs, different temporal responses of PDs and subsequent electronic amplifiers, and different intensities of two output beams, make the positive and negative pulses of two arms unable to be completely eliminated, and the remaining difference will change with the LO fluctuation.
	
	Thus, as shown in Figure~\ref{var_model}, considering the power fluctuation of the practical laser and the imperfection of the beam splitter and PDs, the power fluctuation cannot be removed. Then the measurement noise variance defined as $\sigma_{M}^{2}$ is no longer pure vacuum fluctuation variance, it will be
	\begin{equation}
	\sigma_{M}^{2}=\sigma_{L O}^{\prime 2}+\sigma_{Q}^{\prime 2}+\sigma_{E}^{2},
	\label{eq4}
	\end{equation}
	where $\sigma_{L O}^{\prime 2}$, $\sigma_{Q}^{\prime 2}$, and $\sigma_{E}^{2}$ are the LO fluctuation noise variance, vacuum quantum noise variance, and electronic noise variance, respectively. Only when the detector works in a linear region are the electronic noise and LO intensity independent~\cite{Lvovsky09}, and then the above three variances are independent and the measurement noise variance is the sum of the three.

	Next, we analyze the influence of the LO fluctuation on the conditional min-entropy of the vacuum-based CV-QRNG in the situation of imbalanced homodyne detection in detail. Since the electronic noise variance can generally be regarded as a constant, for convenience we consider it as 0 and leave it to be discuss later when we analyze the measurement noise variance $\sigma_{M}^{2}$ under non-ideal conditions. $\sigma_{M}^{2}$ is derived based on the principle of homodyne detection below.
	\subsection{\label{3.2}Impacts of LO fluctuation under imbalanced homodyne detection}
	The homodyne detection technique is a highly sensitive detection technique, which is sensitive to the amplitude and phase of the input signal. It is used to measure the amplitude and phase of the weak signal beam. When the homodyne detection is unbalanced, the LO fluctuation will not be canceled. The impact of the local oscillator’s noise in the case of unbalanced homodyne detection, which has been studied in the context of pulsed optical-field statistics measurements~\cite{raymer1995ultrafast}and the hacking of continuous-variable quantum key distribution~\cite{ma2013wavelength}, should be taken into account in a practical vacuum-based CV-QRNG.
	
	Figure~\ref{setup_basic} shows the schematic of the homodyne detector (HD) in the dashed box. As shown in Figure~\ref{setup_basic}, the cw laser beam serving as the LO enters the fiber beam splitter. The other input is a vacuum state. Their interference is output to the detector, the two currents obtained are subtracted and amplified subsequently.
	
	In this process, mixed in the beam splitter, the electric fields of signals and the LO are $E_{S}(t)=E_{S}+\delta X_{S}(t)+i \delta P_{S}(t)$ and $E_{L}(t)=\left[E_{L}+\delta X_{L}(t)+i \delta P_{L}(t)\right] e^{i \varphi}$, respectively,
	where $E_{S}$ and $E_{L}$ are real time-independent terms and $E_{L}>>E_{S}$,  $E_{S(L)}+\delta X_{S(L)}(t)$, and $\delta P_{S(L)}(t)$ are real and describe changes of amplitude and phase quadrature of the signal (LO) field. The output electric fields become $\begin{bmatrix}E_{1}\\ E_{2}\end{bmatrix}=\begin{bmatrix} \sqrt{t_{13}} & \sqrt{r_{23}} \\ \sqrt{t_{14}}  & -\sqrt{t_{24}} \end{bmatrix}\begin{bmatrix}E_{L}\left ( t \right ) \\ E_{S}\left ( t \right ) \end{bmatrix}$, where $t_{13}$, $t_{24}$, $r_{23}$ and $r_{14}$ are the corresponding reflection and transmission coefficients of the beam splitter and as shown in Figure~\ref{setup_basic}, with ports 1 and 2 as input ports and ports 3 and 4 as output ports. Then the outputs are detected by the PDs, considering further that the conversion efficiencies of the PDs of the HD are different, named $\eta_{1}$ and $\eta_{2}$, respectively, the difference of the two output currents is
	\begin{equation}
	\label{eq5}
	\begin{split}
	i_{\mathrm{sub}}(t)&=\eta_{1}|E_{1}|^{2}-\eta_{2}|E_{2}|^{2}\\
	&\approx(\eta_{1}t_{13}-\eta_{2}r_{14})[|E_{L}|^{2}+2 E_{L} \delta X_{L}(t)]\\
	&+ 2 E_{L}(\eta_{1}\sqrt{t_{13} r_{23}}+\eta_{2}\sqrt{t_{24} r_{14}})[(E_{S}+\delta X_{S}(t)) \cos \varphi\\
	&+\delta P_{S}(t) \sin \varphi],
	\end{split}
	\end{equation}
	its variance is
	\begin{equation}
	\label{eq6}
	\begin{split}
	\langle i_{\mathrm{sub}}^{2}(t)\rangle &\approx 4 E_{L}^{2}\{(\eta_{1} t_{13}-\eta_{2} r_{14})^{2}[\langle\delta X_{L}^{2}(t)\rangle-\langle\delta X_{L}(t)\rangle^{2}]\\&+ (\eta_{1} \sqrt{t_{13} r_{23}}+\eta_{2} \sqrt{t_{24} r_{14}})^{2}[\delta X_{S}^{2}(t) \cos ^{2} \varphi\\&+\delta P_{S}^{2}(t) \sin ^{2} \varphi] \}.
	\end{split}
	\end{equation}
	With consideration of the trans-impedance gain $g$ of the detector and regarding the electric field of the vacuum state as the signal, the total measurement noise variance is
	\begin{equation}
	\sigma_{M}^{2} = 4g^{2}\rm P\left(a \sigma_{\mathrm{LO}}^{2}+b \sigma_{\mathrm{Q}}^{2}\right),
	\label{eq7}
	\end{equation}
	where $P$ is the optical power, $g$ is the trans-impedance gain of the detector with unit of V/A, and $a=\left(\eta_{1} t_{13}-\eta_{2} r_{14}\right)^{2}$ and $b=\left(\eta_{1} \sqrt{t_{13} r_{23}}+\eta_{2} \sqrt{t_{24} r_{14}}\right)^{2}$ are imbalance coefficients, which are determined by the transmittance and reflectance of the beam splitter and the photoelectric conversion efficiencies of PDs. $\sigma_{Q}^{2}$ is the vacuum quantum noise variance. If $\varphi = 0$, $\sigma_{Q}^{2} = \delta X_{S}^{2}(t)$, $\sigma_{Q}^{2}$ comes from measuring the $X$ quadrature; if $\varphi = \pi/2$, $\sigma_{Q}^{2} = \delta P_{S}^{2}(t)$, $\sigma_{Q}^{2}$ comes from measuring the $P$ quadrature.
	$\delta X_{L}(t)$ is the change of LO amplitude quadrature, and $\sigma_{L O}^{2}=\left\langle\delta X_{L}^{2}(t)\right\rangle-\left\langle\delta X_{L}(t)\right\rangle^{2}$ is the variance of the LO amplitude change.
	
	In consideration of the impacts of the LO fluctuation, the imbalanced homodyne detection, the electronic noise, the discrete variables, and the quantization error, the practical conditional min-entropy is
	\begin{equation}
	\label{eq8}
	\begin{split}
	H_{\min }(M_{d i s} | E)&=-\log _{2}\left[\operatorname{erf}(\frac{\delta}{2 \sqrt{2 \sigma_{Q}^{\prime 2}}})\right]\\&=-\log _{2}\left[\operatorname{erf}(\frac{R}{2^{n+1} \sqrt{2 \sigma_{Q}^{\prime 2}}})\right],
	\end{split}
	\end{equation}
	where $\sigma_{Q}^{\prime2}=\sigma_{M}^{2}-\sigma_{E}^{2}-\sigma_{\mathrm{LO}}^{\prime 2}-3(\delta / 12)^{2}$ and $(\delta / 12)^{2}$ is the quantization error variance.
	
	Though there is a positive correlation between $H_{\min }\left(M_{dis} | E\right)$ and $\sigma_{Q}^{2}$, $\sigma_{Q}^{2}$ cannot be measured directly. We transform the analysis of the conditional min-entropy into the analysis of $\sigma_{M}^{2}$. By taking the electronic noise variance into account, Eq. (~\ref{eq7}) becomes
	\begin{equation}
	\sigma_{M}^{2}=4g^2P\left(a \sigma_{\mathrm{LO}}^{2}+b \sigma_{\mathrm{Q}}^{2}\right)+\sigma_{E}^{2}.
	\label{eq9}
	\end{equation}
	
	This is the same form as Eq. (~\ref{eq4}) in Sec.~\ref{3.1} where $\sigma_{L O}^{\prime 2} = 4ag^2P\sigma_{LO}^{2}$ and $\sigma_{Q}^{\prime 2} = 4bg^2P\sigma_{Q}^{2}$ are the amplified variances of the LO fluctuation and the vacuum fluctuation. Obviously $a \leq 1$, and $b \leq 1$. When $b = 1$, the ideal scenario is that the couple ratio of the beam splitter is 50:50 and the efficiencies of PDs are $100 \%$, thus $a = 0$; i.e., the impact of the LO fluctuation is cleared up. However, there are defects in practical devices, for example, the reflection and transmission coefficients of the beam splitter are less than $50 \%$ and it is impossible for the efficiencies of PDs to reach $100 \%$ due to the techniques and losses, so $b<1$. If only the laser is ideal, that is, there is no LO fluctuation, then even if the beam splitter and PDs are not ideal symmetrical, the total measurement noise after eliminating the electronic noise is complete vacuum quantum noise. In general, the electronic noise of the system obeys a Gaussian distribution, and its variance is invariable. The vacuum quantum noise can be regarded as obeying a Gaussian distribution of $N$ (0,1), and its variance can also be treated as a constant. Furthermore, the values of $a$ and $b$ depend on the imbalance of the beam splitter and the photoelectric conversion efficiencies of PDs.
	
	\begin{figure}[t]
		\centering
		\includegraphics[width=9cm]{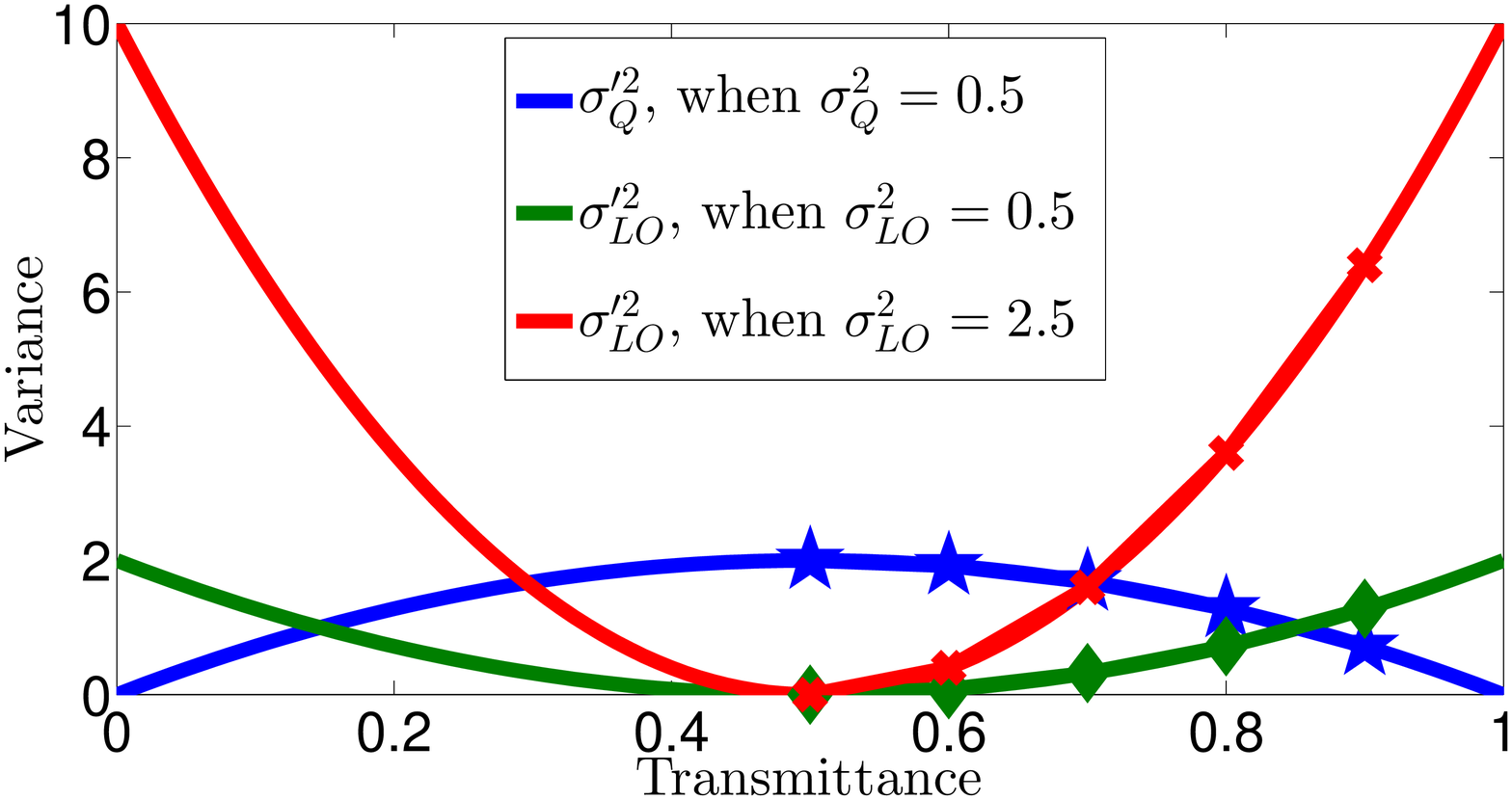}
		\caption{Numerical simulations of $\sigma_{L O}^{\prime 2}$ and $\sigma_{Q}^{\prime 2}$ vs transmittance. Suppose in an ideal scenario, $\eta _1 = \eta _2 = 1$, $a+b=1$, $g = 1$, $P = 1$, $\sigma _Q^2 = 0.5$, and $\sigma _E^2 = 0 $ }
		\label{t_vs_var_lo_q}
	\end{figure}
	
	\begin{figure}[t]
		\centering
		\includegraphics[width=9cm]{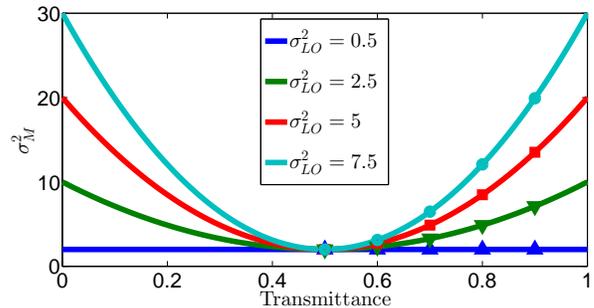}
		\caption{Numerical simulations of $\sigma_{M}^2$ vs transmittance. Suppose, in an ideal scenario, $\eta _1 = \eta _2 = 1$, $a+b=1$, $g = 1$, $P = 1$, $\sigma _Q^2 = 0.5$, and $\sigma _E^2 = 0 $ }
		\label{t_vs_var_m}
	\end{figure}

	\begin{table}[b]
		\caption{\label{tab1}%
			The actual transmittance and reflectance of different couple ratios of the beam splitter with port 1 and port 2 as the input ports and port 3 and port 4 as the output ports.}
		\begin{ruledtabular}
			\begin{tabular}{ccccc}
				\textrm{Couple ratio}&
				\textrm{$t_{13}$}&
				\textrm{$r_{14}$}&
				\textrm{$r_{23}$}&
				\textrm{$t_{24}$}\\
				\colrule
				50/50   & $48.78 \%$    & $47.71 \%$    & $48.93 \%$    & $48.52 \%$\\
				60/40	& $61.25 \%$    & $38.26 \%$    & $38.44 \%$    & $61.38 \%$\\
				70/30	& $69.82 \%$    & $30.17 \%$    & $28.17 \%$    & $63.49 \%$\\
			\end{tabular}
		\end{ruledtabular}
	\end{table}
	
	\begin{figure*}[t]
		\centering
		\includegraphics[width=16cm]{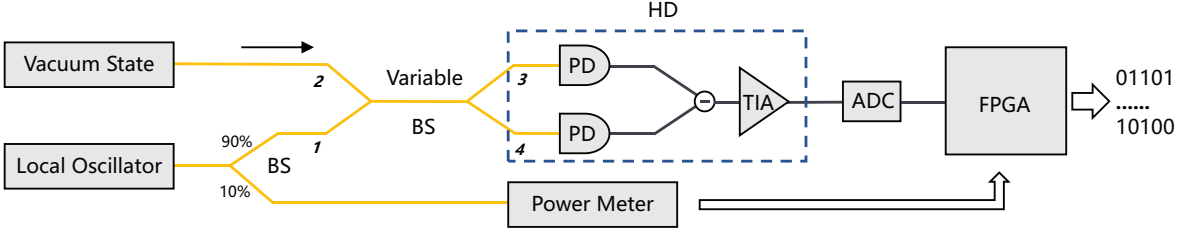}
		\caption{Experimental setup of a vacuum-based CV-QRNG. Using a 90:10 fiber beam splitter, we separate a part of the LO and use the fiber optic power meter (the measurement accuracy is $0.01um$) to monitor and calculate the noise variance of the LO fluctuation, which will be used to calibrate the conditional min-entropy according to Eq. (~\ref{eq8}). The rest of the LO, which interferes with the vacuum state in the fiber beam splitter, is detected by the homodyne detector and followed by sampling and postprocessing operations to yield random numbers. }
		\label{setup_monitoring}
	\end{figure*}

	\begin{figure}[b]
		\centering
		\includegraphics[width=9 cm]{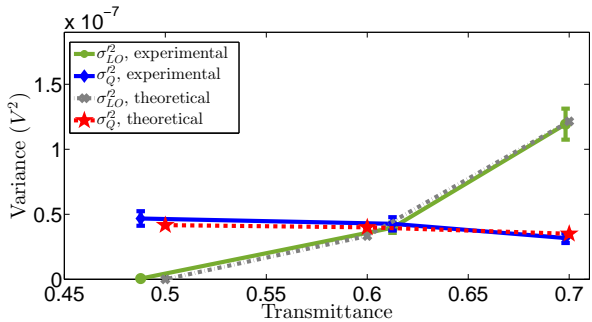}
		\caption{$\sigma_{LO}^{\prime 2}$ and $\sigma_{Q}^{\prime 2}$ vs transmittances for different beam splitters when $P=1.066$ mW.
			The solid lines are experimental results with error-bars that are measured and calculated in practical scenarios. The dash lines are theoretical results that are calculated in the ideal case of perfect beam splitters and a pure vacuum state (i.e. $\sigma_{Q}^2=0.5$). }
		\label{cr_vs_lo_q_sim}
	\end{figure}
	
	\begin{figure}[b]
		\centering
		\includegraphics[width=9 cm]{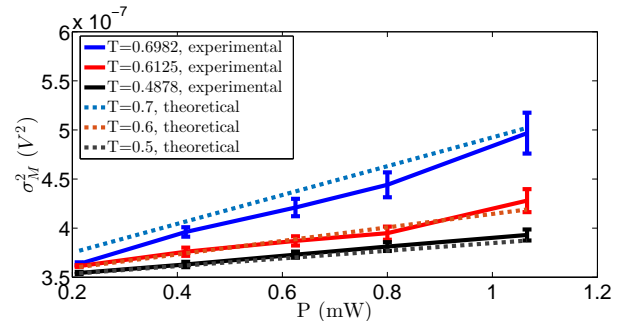}
		\caption{Practical $\sigma_{M}^{2}$ vs optical power in different beam splitters (transmittances are 0.4878, 0.6125 and 0.6982, from bottom to top) with error bars.
			To get a better comparison of the results of such transmittances, the appropriate power is set as 1.066 mW, because results for transmittances of $0.6982$ and $0.6125$ will be saturated over 1.066 mW successively.}
		\label{p_vs_m_sim}
	\end{figure}
	
	According to Eq.~(\ref{eq9}), we suppose in an ideal scenario, $\eta _1 = \eta _2 = 1$, $a+b=1$, $g = 1$, $P = 1$, $\sigma _Q^2 = 0.5$, $\sigma _E^2 = 0 $, and $\sigma _{LO}^2 = \sigma _Q^2,5\sigma _Q^2,10\sigma _Q^2, and 15\sigma _Q^2$ (estimated from our measurement results that $\sigma _{LO}^2 / \sigma _Q^2$  is about 14.78$\sim$18.94). For convenience, transmittance is used to describe the unbalance of the beam splitter, and the larger the transmittance is, the greater the unbalance is. Numerical simulations are done to reveal the analytic relations between the transmittance of the beam splitter and the variances ($\sigma_{L O}^{\prime 2}$, $\sigma_{Q}^{\prime 2}$ and $\sigma_{M}^2$), which are shown in Figures~\ref{t_vs_var_lo_q} and~\ref{t_vs_var_m}. In the figures, the dots represent when the transmittance equals 0.5, 0.6, 0.7, 0.8, and 0.9, respectively and the corresponding $a$ is 0, 0.04, 0.16, 0.36, and 0.64 and $b$ is 1, 0.96, 0.84, 0.64, and 0.36. We can cautiously draw inferences from Figures~\ref{t_vs_var_lo_q} and~\ref{t_vs_var_m} that the greater the transmittance of the beam splitter is, the larger $\sigma_{L O}^{\prime 2}$ is, the larger $\sigma_{M}^2$ is, and the smaller $\sigma_{Q}^{\prime 2}$ is with certain trans-impedance gains $g$, and power $P$. For instance, when $\sigma_{L O}^2 = 5\sigma_{Q}^2 = 2.5$ and transmittance increases by $20\%$ from 0.5, $\sigma_{L O}^{\prime 2}$ will increase to 0.4 from 0, $\sigma_{Q}^{\prime 2}$ will reduce by $4\%$ from 2, and $\sigma_{M}^2$ will increase by $16\%$ from 2. In particular, while $\sigma_{L O}^2 = \sigma_{Q}^2$, $\sigma_{M}^2$ will remain unchanged.
	\section{\label{set4}Experimental Demonstration}
	
	To further study the impacts of the imperfect factors discussed above on the measurement results and the practical conditional min-entropy, we set up the experimental system as shown in Figure~\ref{setup_monitoring} shown. Firstly, the performance of the practical devices of the vacuum-based CV-QRNG system is measured. Among the devices, the 12-bit ADC is homemade, and the Xilinx KC705 evaluation board is used as the FPGA platform. The conversion efficiencies of two PDs in the balanced amplified photodetector (THORLABS, PDB450C, gain adjustable) are measured as follows: $\eta_{1}=0.584 A/W$ and $\eta_{2}=0.561 A/W$. The actual transmittances and reflectances of the three selected beam splitters with different couple ratios are shown in Table I, from which the values of $a$ and $b$ can be calculated. While the LO is input from port 1, the practical transmittance is measured as $t_{13}=P_{port3}/P_{port1}$.
		
	Because the imbalance of the beam splitter easily leads the detector to saturation, it is difficult to improve the optical power range and it is not conducive to observe the effect of the LO fluctuation on $\sigma_{M}^{2}$ in different optical power. To avoid this, we choose the detector with the lowest trans-impedance gain (1kV/A) and the bandwidth of dc of 150MHz. In order to obtain the practical conditional min-entropy, we propose a scheme to monitor the LO and calibrate the practical conditional min-entropy, which is similar to the method in continuous-variable quantum key distribution system~\cite{zhang2019continuous,zhang2020one} that uses the $10\%$ of the LO for clock synchronization, data synchronization, and LO monitoring. As shown in Figure~\ref{setup_monitoring}, using a 90:10 fiber beam splitter, we separate part of the LO and use the fiber optic power meter (ILX Lightwave FPM-8210H) to monitor and calculate the noise variance of the LO fluctuation.
	For instance, the “$10\%$ beam” (monitoring path) is measured by the power meter and its power fluctuation is named $\sigma_{mon}^2$. For an ideal 90:10 beam splitter, the “$90\%$  beam” is 9 times that of the $10\%$ one, thus the LO fluctuation variance could be calculated as $\sigma_{LO}^2=9\times\sigma_{mon}^2$. While for the practical one in our setup, the multiple is 9.85 and $\sigma_{LO}^2=9.85\times\sigma_{mon}^2$.

	At the beginning, by using three beam splitters of different transmittances (0.4878, 0.6125 and 0.6982) in the practical setup, the LO fluctuation variances $\sigma_{LO}^{2}$ and the total measurement variances $\sigma_{M}^{2}$ are measured at the optical power of 1.066 mW. In succession, $\sigma_{L O}^{\prime 2}$ and $\sigma_{Q}^{\prime 2}$ are calculated based on these measurement results. As shown in Figure~\ref{cr_vs_lo_q_sim}, in consideration of statistical fluctuations of the ten measurements on each transmittance, the experimental results of $\sigma_{L O}^{\prime 2}$ and $\sigma_{Q}^{\prime 2}$ are displayed as solid lines with error-bars. It can be seen from Figure~\ref{cr_vs_lo_q_sim} that at the same optical power, the greater the transmittance is, the larger $\sigma_{L O}^{\prime 2}$ is and the smaller $\sigma_{Q}^{\prime 2}$ is. In the meantime, the error gets larger with the increase of the transmittance. Moreover, we calculate $\sigma_{L O}^{\prime 2}$ and $\sigma_{Q}^{\prime 2}$ in ideal circumstances with a pure vacuum state ($\sigma_{Q}^2=0.5$~\cite{Marangon17}) and perfect beam splitters which have accurate transmittances such as 0.5 and 0.6. The theoretical results are plotted as dashed lines in Figure~\ref{cr_vs_lo_q_sim} for comparison. It can be seen that the experimental results are close to the theoretical results but differ from the theoretical results due to practical imperfections. We also calculate the practical conditional min-entropy of the experimental results according to Eq. (~\ref{eq8}). When practical transmittances are 0.4878, 0.6125 and 0.6982 respectively, the practical conditional min-entropies will be 1.15, 1.08, and 0.87 bits. Meanwhile, the ratios of  $\sigma_{L O}^{\prime 2} : \sigma_{Q}^{\prime 2}$ are $1.22\%$, $92.94\%$ and $376.39\%$; hence we can find out that the larger the transmittance is, the greater the impact caused by the LO fluctuation on the practical conditional min-entropy will be.

\begin{figure}[t]
		\centering
		\includegraphics[width=9 cm]{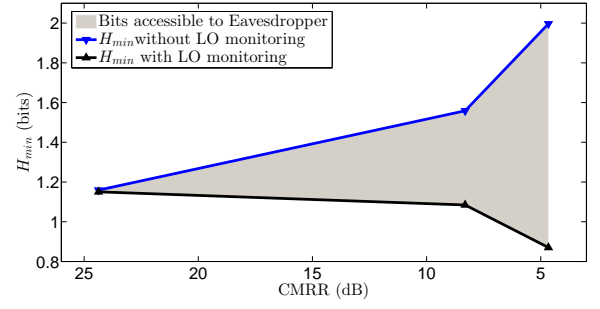}
		\caption{Min-entropy vs CMRR in the cases with or without LO monitoring. The blue line is the min-entropy without LO monitoring. The black line is the min-entropy with LO monitoring. The gray area is bits which are accessible to the eavesdropper.}
		\label{m-e_vs_q_cmrr}
	\end{figure}

	Then the total measurement noise variances $\sigma_{M}^{2}$ with three beam splitters of different transmittances (0.4878, 0.6125 and 0.6982) are measured at the different optical powers. In consideration of statistical fluctuations of the ten measurements at each power, the experimental results $\sigma_{M}^{2}$ are presented as solid lines with error-bars in Figure~\ref{p_vs_m_sim}. Similar to Figure~\ref{cr_vs_lo_q_sim}, the dashed lines indicate the theoretical model that is calculated from Eq.~(\ref{eq9}) by supposing perfect beam splitters, a pure vacuum state and
	the same $\sigma_{LO}^{2}$ (measured at 1.066 mW) at every optical power. As shown in Figure~\ref{p_vs_m_sim}, the total measurement noise variance and its error increases with the incremental power and transmittance. For a certain power, $\sigma_{M}^{2}$ grows with the increment of the transmittance. We note that in a practical system, because of imperfect practical components and the statistical errors, there is a mismatch between the theoretical model and the experimental results.
	The above conclusions drawn from the experimental results are consistent with the analyses of Eq.~(\ref{eq9}) in Sec.~\ref{3.2}.
	\begin{figure}[t]
		\centering
		\includegraphics[width=9 cm]{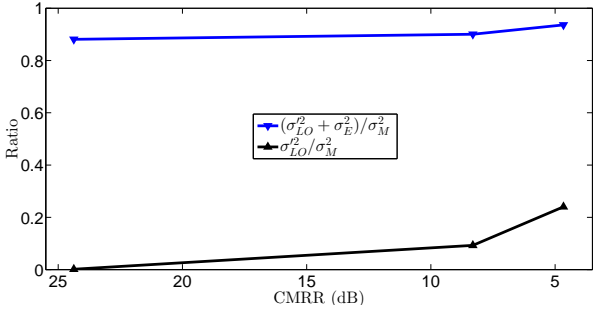}
		\caption{Ratio of classical noise variance to  $\sigma _M^2$ vs the CMRR. The blue line is the proportion of classical noise in the total measurement noise. The black line is the proportion of LO noise in the total measurement noise.}
		\label{loebm_vs_cmrr}
	\end{figure}
	From an adversarial perspective, classical parameters in the system such as the couple ratio of the BS, and the efficiency of the PDs could be accessible and result in an undetectable bias of the output, which would eventually change the common mode rejection ratio (CMRR) of a homodyne detector~\cite{chi2011balanced}. Thus CMRR could be a more general and comprehensive index for the unbalance of homodyne detection. As shown in Figure~\ref{loebm_vs_cmrr}, we experimentally demonstrate the relation between the amount of classical noise in the total measurement noise and the CMRR of the homodyne detector (three different CMRR values are corresponded with three BSs with ratios of 50:50, 60:40, and 70:30). We find that, as the CMRR of the homodyne detector decreases, the amount of classical (LO) noise presented in the measurement of the quantum state increases. Since the amount of LO noise in the total measurement noise increases, an eavesdropper will be more accessible to the side information. Thus we should monitor the LO and calibrate the min-entropy. As shown in Figure~\ref{m-e_vs_q_cmrr}, we also experimentally demonstrate the merit of our proposed security solution. It can be seen in Figure~\ref{m-e_vs_q_cmrr} that as the CMRR of a homodyne detector decreases, an eavesdropper will obtain more bits (shown as the gray area). In the meantime, one can discard these insecure bits with LO monitoring.
	
	\begin{figure}[b]
		\centering
		\includegraphics[width=10 cm]{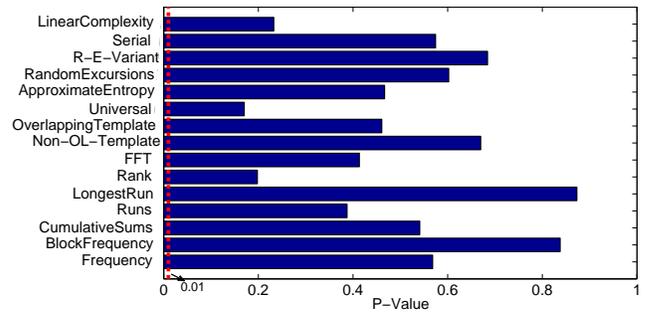}
		\caption{Results of the NIST statistical test suite (15 test items are contained). If the $P$ value satisfies $0.01\le$ $P$ value$\le 0.99 $, the test is considered successful. The dashed line indicates $P$ value$=0.01$. }
		\label{nist_result}
	\end{figure}
	
	Finally, we demonstrate an experiment to obtain the practical conditional min-entropy of a practical system. We find an optimal power by carefully increasing the power from $0$ mW. When the power is $2.26$ mW, the total measurement noise variance $\sigma _M^2  = 4.26 \times {10^{ - 7}}{V^2}$, the electronic noise variance  $\sigma _E^2  = 3.47 \times {10^{ - 7}}{V^2}$,and the noise variance of the LO fluctuation is  measured as $\sigma _{LO}^2=1.21 \times {10^{ - 9}}{V^2}$. Then the measurement results of  $\sigma_{M}^{2}$,  $\sigma_{E}^{2}$  and  $\sigma_{LO}^{\prime 2}$  are substituted into Eq. (~\ref{eq8}) with a quantization error variance of $1.21 \times {10^{ - 9}}{V^2}$, and the practical conditional min-entropy is ${H_{\min }}\left( {{M_{{\rm{dis}}}}{\rm{|}}E} \right){\rm{ = }}1.40$ bits. Though it is clear that $\sigma_{LO}^{\prime 2}$ is at the same order as the quantization error variance, a tighter bound of the conditional min-entropy should take it into account.

	For a 12-bit ADC, the extraction ratio of a Toeplitz matrix can be set as $10\%$ and the dimensions of the matrix are $768\times7680$ with the information theoretic security parameter $\varepsilon {\rm{ =  5}}{\rm{.42}}{{\rm{e}}^{{\rm{ - 20}}}}$.
	With a detection bandwidth of dc of 150MHz and a sampling rate of 300MHz, the corresponding random-number generation rate is calculated as $360 Mbps$. The generated random-numbers are passed through all 15 tests of the NIST test suite, the test results are shown in Figure~\ref{nist_result}.

	\section{\label{set5}Conclusion and Discussion}
	In this paper, we discuss the LO fluctuation under imbalanced homodyne detection due to imperfections of practical devices, which will influence on the practical security of a vacuum-based CV-QRNG. Based on the theory of the conditional min-entropy and the positive correlation between the conditional min-entropy and the variance of the quantum noise, effects from imperfections of practical devices on the practical security of a vacuum-based CV-QRNG are analyzed in detail, such as the LO fluctuation of the laser, the imbalance of the beam splitter and the limited conversion efficiencies of the PDs. In particular, we develop a practical variance model and a practical conditional min-entropy model.
	Through formula derivation and simulation analyses, we preliminarily find that with certain trans-impedance gain and power, the greater the transmittance is, the larger the amplified LO fluctuation variance $\sigma_{L O}^{\prime 2}$ is, the smaller the amplified quantum noise variance  $\sigma_{Q}^{\prime 2}$ is, and the greater the total measurement variance $\sigma_{M}^2$ for the rapidly-rising $a$ is. Then the experiments we demonstrated prove the theoretical inferences. Moreover, we find that the errors of the experimental results get larger with the increase of the transmittance. In the meantime, experimental results and theoretical results of $\sigma_{L O}^{\prime 2}$ and $\sigma_{Q}^{\prime 2}$ are compared, and it is shown that the experimental results are close to the theoretical results but differ from the theoretical results due to practical imperfections and the statistical errors. Besides, it could be seen from an adversarial perspective that in a practical system, as the CMRR of a homodyne detector decreases, which means the unbalance increases, then an eavesdropper will obtain more bits.
	In the experiments, we propose a LO monitoring method to monitor and calculate $\sigma_{L O}^{\prime 2}$ and at last calibrate the practical conditional min-entropy of a vacuum-based CV-QRNG. Finally, we achieve a generation speed of over $350 Mbps$ for a practical vacuum-based CV-QRNG with a DC coupling HD and LO monitoring.
	
	Recently, a continuous-variable eavesdropping attack, based on heterodyne detection, on a trusted quantum random-number generator has been realized experimentally, which discusses a source of side information for eavesdroppers while additional classical noise beyond the quantum limit~\cite{Thewes19}. However, like other frameworks of practical vacuum-based CV-QRNG, they all have a basic assumption that the LO power is stable. Therefore, it is worth mentioning that, for either practical vacuum-based CV-QRNGs or semi-device-independent QRNGs based on a vacuum state, our work will be complementary to their analysis framework.

	In addition to the influence of the LO fluctuation under imbalanced homodyne detection on the practical security of a vacuum-based CV-QRNG discussed in this paper, there are other interesting issues worthy of further study based on our work. For example, input beams with different wavelengths will lead to varied couple ratios of the beam splitter and different responsivities of PDs, which will affect the security of the practical vacuum-based CV-QRNG. There might attacks against these defects, such as a wavelength attack of wavelength-adjustable laser or changing the couple ratio by controlling the temperature of the beam splitter. Some measurement experiments on beam splitter transmittance at different wavelengths have been done to study the wavelength attack against continuous-variable quantum key distribution~\cite{ma2013wavelength, Li11, Huang14}, which are sound references to the issue in vacuum-based CV-QRNGs.

	
	\begin{acknowledgments}
		This work was supported by the Key Program of the National Natural Science Foundation of China under Grant No. 61531003, the National Natural Science Foundation under Grant No. 61427813, the Fund of CETC under Grant No. 6141B08231115, and the Fund of State Key Laboratory of Information Photonics and Optical Communications.
	\end{acknowledgments}
	
	\bibliography{ref0}
	
\end{document}